Membrane proteins and proteomics: love is possible, but so difficult.

Thierry Rabilloud 1,2

1 CEA-DSV/iRTSV/LBBSI, Biophysique et Biochimie des Systèmes Intégrés, CEA-Grenoble, 17 rue des martyrs, F-38054 GRENOBLE CEDEX 9, France

2 CNRS UMR 5092, Biophysique et Biochimie des Systèmes Intégrés, CEA-Grenoble, 17 rue des martyrs, F-38054 GRENOBLE CEDEX 9, France

Correspondence :

Thierry Rabilloud, iRTSV/BBSI

CEA-Grenoble, 17 rue des martyrs,

F-38054 GRENOBLE CEDEX 9

Tel (33)-4-38-78-32-12

Fax (33)-4-38-78-44-99

e-mail: Thierry.Rabilloud@ cea.fr

Abstract
Despite decades of extensive research, the large-scale analysis of membrane proteins remains a difficult task. This is due to the fact that membrane proteins require a carefully balanced hydrophilic and lipophilic environment, which optimum varies with different proteins, while most protein chemistry methods work mainly, if not only, in water-based media.
Taking this review (Santoni, Molloy and Rabilloud, Membrane proteins and proteomics: un amour impossible? Electrophoresis 2000. 21,1054-1070) as a pivotal paper, the current paper analyzes how the field of membrane proteomics exacerbated the trend in proteomics, i.e. developing alternate methods to the historical two-dimensional electrophoresis, and thus putting more and more pressure on the mass spectrometry side. However, in the case of membrane proteins, the incentive in doing so is due to the poor solubility of membrane proteins.
This review also shows that in some situations, where this solubility problem is less acute, two-dimensional electrophoresis remains a method of choice. Last but not least, this review also critically examines the alternate approaches that have been used for the proteomic analysis of membrane proteins

1. The happy protoproteomics days

In any cell, membrane proteins are part of the interface between the outside and the inside of the cell. As such, they are implicated in key cellular functions such as small molecules transport, cell-cell and/or cell-pathogen and/or cell substrate recognition and interaction, cell communication and signalling, with all the possible modifications in cell functionality that can arise subsequently to the activity of these membrane proteins. Consequently, it is no surprise that from its very beginning, i.e. in the paleoproteomics times (1974-1993), proteomics as a discipline has shown recurrent and continued interest in the analysis of the particular protein subset represented by membrane proteins.
As early as 1976, the first paper dealing with a proteomic analysis of membrane proteins was [1]. At that time however, the protein identification means could not be coupled with the gel electrophoretic techniques, so that it was easy to claim for an improved analysis of membrane proteins by changing the experimental conditions and showing just more protein spots on the 2D gel. The situation went on these lines for a few years (e.g. in [2], until the first protein characterization tools compatible with 2D electrophiresis cam, at the end of the 70's: silver staining [3] and protein blotting [4].
With those tools, it was possible to perform two types of experiments. Either purifying a membrane protein and then analyzing it by 2D electrophoresis [5] [6], or analyzing a membrane preparation by 2D electrophoresis, blotting to the adequate support and identifying it by immunodetection [7] [8]. However, these methods were very of limited scope (e.g. because of the requirement for an appropriate antibody), so that our vision of the real performances of 2D gels for the analysis of membrane proteins was still very fragmentary and thus optimistic. At that times, it was still possible to claim for a better analysis of membrane proteins by just altering the experimental conditions and detecting more protein spots. Tests were carried out mainly on detergents, which are widely used to solubilize membrane proteins, and various detergents such as CHAPS [9], linear sulfobetaines [10], dodecyl maltoside [11]. and amido sulfobetaines [12] were tested for improved solubilization of proteins in membrane preparations.

2. Switch on the light, and face the situation

At the end of the 80's, important improvements appeared that triggered the switch into the real proteomics age. These were on the one end the maturation of immobilized pH gradients for the first dimension of 2D PAGE [13], bringing with them the ability to load enormous amounts of proteins [14], [15], and the other hand the ability to identify any sufficiently abundant spot by protein chemistry techniques. This was achieved first by Edman N-terminal microsequencing from blots [16], then by internal Edman sequencing after peptide separation [17], and finally by mass spectrometry [18], [19], [20], [21]. With such an arsenal in hands, it was finally possible to knows the real performances of 2D electrophoresis in the analysis of membrane proteins, and it took a few years to figure out the situation: 2D gels had enormous difficulties to display membrane proteins [22], [23]. But difficult does not means impossible, and it was soon shown that the flexibility of IPGs was very helpful. By changing the chaotrope [24] and the detergent [25], [26], bona fide membrane proteins (i.e. proteins with predicted transmembrane segments) could be displayed on 2D gels.

It was at that time (difficult but not impossible) that the review rooting this paper [27], was written and published. It showed exactly that there were a few examples of membrane proteins analyzed by 2D gel electrophoresis, with one happy exception,  that was bacterial membranes, and especially the outer membrane of Gram- bacteria. It also showed that even when a membrane protein could be detected on a 2D gel, this did not mean that it was efficiently and properly extracted. This variety in performance of solubilization by 2D gels is shown on Figure 1.
In the past 8 years, this paper has attracted more than 450 citations, which means that many people found it useful. In fact, this paper showed that the analysis of membrane proteins by 2D electrophoresis was as a half-filled glass: half-full but also half-empty. Consequently, the two behaviors were found among the scientists interested in this field.

3. Progress in the performances of 2D gels

Some people found that the glass was half-full, so that by further altering the conditions used in 2D electrophoresis, the situation could be improved. This meant mainly testing new combinations of surfactants and chaotropes on various membrane preparations. these detergents can be various sulfobetaines [28], [29], nonionic detergents [30], dodecyl maltoside [31], or short chain lecithins [32]. These papers showed indeed that it was possible to solubilize more and more membrane proteins [33]. It was also shown that a detection problem may arise on 2D gels in the specific case of membrane proteins [34]. It was also shown that even that peptides corresponding to transmembrane segments can be visualized by the MS analysis of 2D gel-separated membrane proteins [35]. If mitochondria were a good example for the success of this approach, even with studies not targeting membrane proteins per se [36], an even better example is represented by bacterial membranes. Following the seminal paper of Molloy and coworkers [37], many researchers used opimized 2D gels to analyze bacterial membrane proteins. While the success has been limited for Gram+ bacteria [38], [39], where there is only one membrane with most proteins spanning the membrane by helices, much more work has been devoted to Gram- bacteria, with greater success such as the identification of a new OMP [40], the study of E. coli either from the basic microbiology point of view [41] or from the side pathogenic bacterial strains [42], [43], and more generally the study of various Gram- organisms of interest in various areas [44] [45] [46] [47] [48] [49] [50] [51] [52]. This happy exception is due to the fact that many proteins in the outer membrane of Gram- bacteria are of the porin type, which means that their transmembrane part is not made of helices but of a beta barrel. Once properly denatured, these proteins are fairly soluble in the conditions prevailing in IEF, and can thus be analyzed with high resolution. this is true for bacterial porins [37] but also for eukaryotic porins [53]. This explains why in this special case of bacterial membranes, 2D gel electrophoresis is still widely and efficiently used, even in very recent publications [42].

4. Pendulum at the other side: analysis without protein electrophoresis

However, despite these successes, most scientists felt that the glass was rather half -empty. when combined, this review [27] and the paper showing that the general hydropathy of the polypeptide chain had a major impact on its ability to be seen in 2D gels [23] convinced many researchers that 2D gels would never be an efficient tool to analyze membrane proteins. An excellent example is found in a publication analyzing erythrocyte membranes [54], demonstrating that there is a specific problem with the IEF dimension of 2D gels. This was suspected from the beginning, but demonstrated later [55], [56] [57].
Thus, if there is a problem with the IEF dimension, what is possible to do.
The most drastic idea is to avoid electrophoresis at all, following the shotgun concept introduced by Yates and coworkers [58]. In this case, the basic idea is that even the worst hydrophobic protein will produce at least one nice hydrophilic peptide upon trypsin digestion, and that this peptide will be detected in the shotgun approach, thereby flagging the presence of the membrane protein (to be precise, the presence of at least a protein fragment containing this peptide. This idea works indeed, as shown in a recent review [59], and it works even better when an optimized digestion with proteinase K is used [60].

However, other strategies can be used. If the IEF dimension is a problem, then the most obvious strategy is to replace IEF by another electrophoretic method in a 2D gel process.

5. An intermediate solution; 2D gels without IEF

In this respect, the contrast between the poor solubilizing performances under IEF conditions and the excellent ones in SDS PAGE gives directions to be followed for dedicated solubilization systems for electrophoretic separation of membrane proteins. Of course, membrane proteins must be separated in the presence of detergents to cover the hydrophobic parts of the protein. But in addition, the electrostatic repulsion between molecules must be maximal to prevent aggregation. To this purpose, it is often advantageous to add extraneous charges to proteins via a charged protein-binding agent. Consequently, two main types of 2D gels can be used for separating membrane proteins:

In the first type, the first dimension uses native electrophoresis of membrane proteins and/or membrane complexes, generally with a charge modifying agent. This concept, which traces back very early in electrophoresis and has been reviewed previously [61], has been refined and further developed more recently, using either Coomassie blue as a charge transfer agent [62], or an ionic detergent as a charge transfer agent, in both cases in conjunction with a nonionic detergent [63], [64]. In this case, the rationale is to separate protein complexes, so that the total hydrophobicity of the complex is averaged between all the subunits. This allows separation of hydrophobic proteins when they are embedded in a complex [65], [66] but there is no documentation of separation of monomeric hydrophobic proteins.

In the second type, the first dimension separates denatured proteins, and in this case, the denaturing agent is most often also the charge transfer agent, and is made of an ionic detergent. Of course, it would be of little interest to use twice the simple SDS-PAGE technique, as the proteins would simply lay on the diagonal of the gel. Thus, the optimal system in the first dimension should offer a separation as different as possible from SDS PAGE, while keeping the high loading capacity and high solubilizing power of SDS PAGE.

Among the various possible electrophoretic systems, the urea-16BAC system originally devised by MacFarlane [67] has most of these desirable features. It shows a high loading capacity [68], while also showing a very different migration when compared to SDS [69]. Its ability to separate bona fide integral membrane proteins and thus its utility in membrane proteomics was demonstrated rather early [70], so that this system became an obvious choice for the people disappointed by classical 2D PAGE for membrane proteins, as shown for example on bacterial membrane proteins [71]. It therefore received a lot of applications in various fields, spanning from bacterial proteins [72], [73], to yeast [74], to mammalian cells or tissues [75], [76], [77] or to subcellular membranes [78].

Thus, this system, as well as closely-related ones using other cationic detergents such as CTAB instead of 16-BAC [79] have gained increased popularity in membrane proteomics. Compared to the initial description [67], the newest versions do not use the staining between the two dimensions, but in a simple equilibration in the SDS-containing buffer [79].

However, the cationic system is not as straightforward to use and not as versatile as SDS PAGE.

For example, the polymerization of gels at low pH cannot be achieved by the classical and robust TEMED/persulfate system, and the more delicate ascorbate/ferrous ion/hydrogen peroxide system is often used. Alternatively, the methylene blue-based photopolymerization system [80], which has been shown to be efficient with other types of acidic, urea-containing gels [81], can be used to polymerize the acidic first-dimension gels [82]. Moreover, in this system, that the solubilization is not fully driven by the cationic detergent, but rather both by the native charge of the proteins and the charges added by the detergent. This suggests in turn that the cationic detergents are less efficient than SDS for the solubilization of proteins, so that it can be expected that some membrane proteins are soluble in SDS-containing media and not in cationic detergents-containing media. This assumption has recently been demonstrated as true [83].

6. Keep on the safe side: SDS-only separations

Thus, if optimal membrane protein solubilization is required, SDS must be used in both dimensions. Consequently, some tricks must be found to reach somewhat different separations in the two SDS-based separations. This can be achieved to some extent by changing the buffer system from one dimension to another, as this alters the resolution of the system [84]. To further enhance this differential mobility effect, nonionic modifyers can be used, such as glycerol [85] or urea [86]. Both approaches have been shown to solubilize adequately membrane proteins [83], [87], and the urea-based approach has been shown to be superior to the cationic detergents-based approaches in terms of hydrophobic proteins solubilization [83]. However, this comparison also showed that the resolution on 2D gels, in terms of spot spreading on the gels, clearly ranges in the order: double SDS techniques < cationic/anionic detergent techniques << IEF-based techniques. While this was easily predictable on the basis of the interdependency of the separation principles used for the two-dimensions of the system, this is not without consequences on how we can use these electrophoretic systems optimally in proteomics studies and on how we can improve their use. In fact, the resolution of the non-IEF 2D systems is usually not sufficient enough to separate the proteins present in complex membrane samples. This means that in many cases, the protein spots present in those "off-diagonal" 2D gels contain several proteins. This means in turn that we cannot use image analysis, as we always do with classical 2D electrophoresis, to determine protein amounts variations from spot volume variations. This means that the intermediate readout represented by image analysis is no longer usable, unless the sample is so simple that its complexity matches with the limited resolution of the off diagonal electrophoreses (e.g. in [83]) and that only the final readout represented by peptides in the mass spectrometer can be used for peptide quantification.

Going along this trend, many scientists disappointed by the performances of 2D gels for the analysis of membrane proteins went directly to another approach, combining a single and simple SDS-PAGE as the protein separation, and a peptide separation after in gel digestion to cope with the protein complexity in each protein band in the gel. This approach has been shown to work on complex samples [88], [89] and was shown to be very efficient on membrane proteins [90]. Combined with isotopic labelling [91], this technique is able to provide a quantitative view of membrane proteins. This applicability to the analysis of membrane proteins has been shown in several systems, ranging from organelles such as the lysosome [92] [93] to bacterial membranes [94], [95], [96] or plant membranes [97], or membranes from animal cells [98], [54], [99], [100].

However, this approach is very sensitive to one key factor, which is the number and relative abundance of proteins for each band cut and analyzed from the SDS gel. If many proteins of widely divergent abundances are present within the same band, it is more likely that the more minor ones will be poorly characterized, if not missed.

7. Room for improvement

The emerging trend after all these years is that classical 2D gel is definitively not the most appropriate approach to analyse membrane proteins with mutiple transmembrane domains. However, this leaves the scientific community short of a high resolution separation method for these proteins, and this means in turn that other separation methods with lower resolutions cannot be used as intermediate readouts in the analysis of complex membrane separations. This means consequently an increasing (over)loading on the mass spectrometer, increasing with the complexity of the samples.

Thus, in the SDS PAGE-base system, but this is also true in the off-diagonal 2D systems, any enrichment strategy decreasing the number of unwanted proteins (e.g. soluble contaminants always present in membrane preparations) will be very interesting to apply, provided that it will not alter the reproducibility of the experiment. The most widely-used enrichment is made by washing membranes in high salt and/or high pH solutions. However, this leads to very limited enrichment in intrinsic membrane proteins, which represent ca. 10 percent of the proteins in the unwashed membranes and ca. 20 percent in the washed ones [101]. Enrichment by two-phase partitioning has also been described, but the real gain in performance has not been quantified [102].

A direct and straightforward approach could be to select the intrinsic membrane proteins on the basis of their hydrophobicity, e.g. by extracting them into organic solvents [103], [104]. Although this method clearly achieves a major enrichment in hydrophobic proteins [104], some more soluble, non membrane proteins are also extracted [103], and the methods has an opposite bias when compared to alkaline washes: while alkaline washes let a lot of non membrane proteins go with the membrane fraction, organic extraction leaves a lot of membrane proteins in the pellet with soluble ones.

This analysis shows that no satisfactory method is currently available to enrich membrane preparations selectively in proteins bearing one or more transmembrane domains, and in the absence of a high resolution separation method able to deal with such proteins, this remains a real bottleneck

8. Concluding remarks

As a conclusion, the 2000 paper [27] was not a breakthrough paper (and a review paper can hardly be so). It was just a timely review, helping people to become conscious of the real situation in membrane proteomics. As such, if this paper promoted some research in the field of membrane proteomics, either to improve classical 2D electrophoresis or to devise alternate methods, it really fulfilled its role as a review paper.

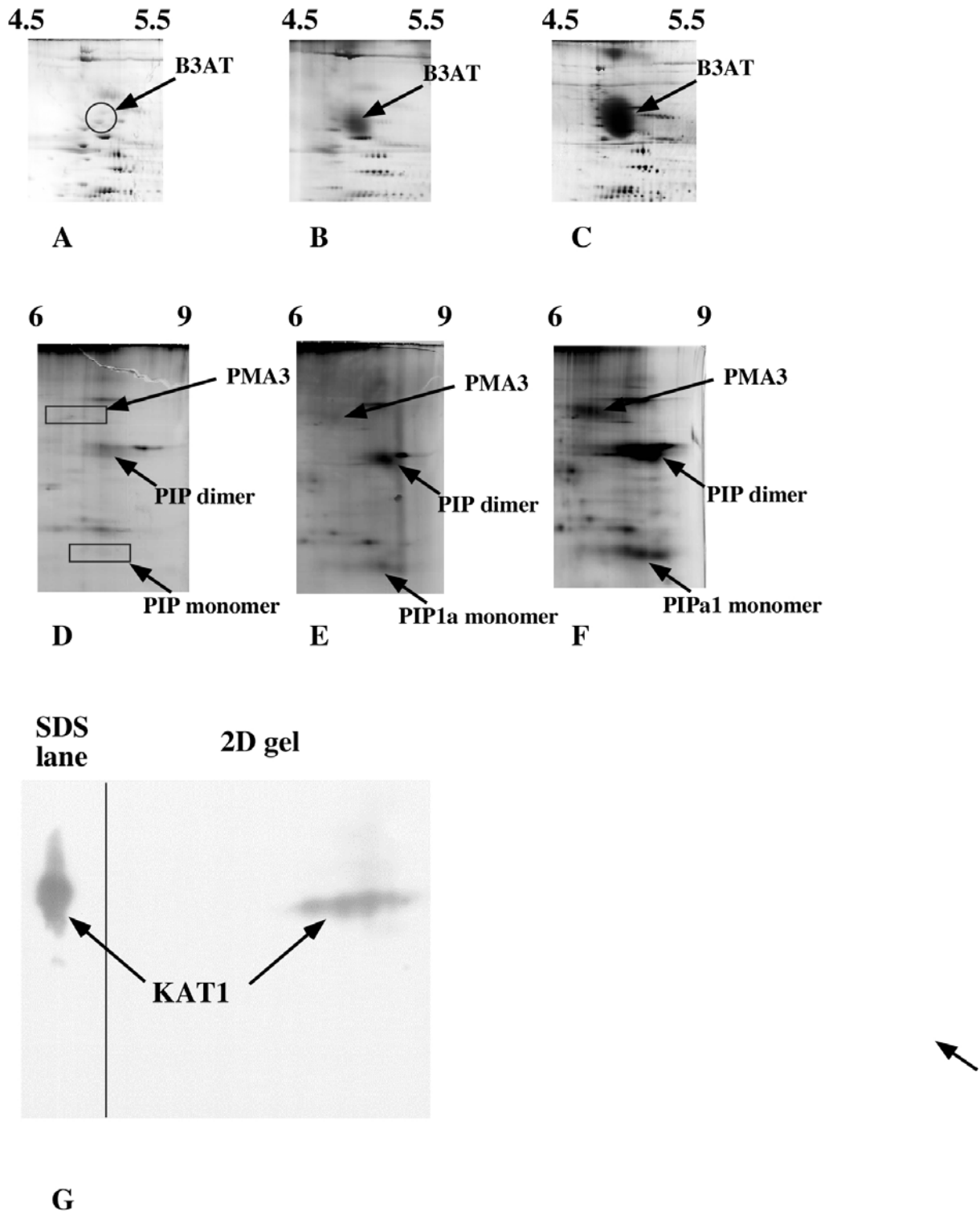

Figure 1: Examples of membrane proteins separation by 2D gels

First row: when it works part 1
Ghosts from human erythrocytes were used as a starting material. They were extracted in urea+thiourea+detergent, and separated by 2D gels (100 µg loaded). Only the zone around the anion transporter B3AT (100kDa, 12 TMDs) is shown.
A: detergent was CHAPS; B detergent was Brij96; C: detergent was dodecyl maltoside
This shows how a change in detergent progressively solubilizes this membrane protein, which appears fuzzy due to its glycosylation

Second row: when it works part 2
A. thaliana leaf plasma membrane proteins were used as a starting material. They were extracted in urea+thiourea+detergent, and separated by 2D gels (60 µg loaded). The basic zone, encompassing the major aquaporin PIP1a (30 kDa, 6TMDs) and the major proton ATPase PMA3 (105kDa, 10 TMDs), is shown
D: detergent was CHAPS; E detergent was ASB14; F: detergent was dodecyl maltoside

G: When it fails.
Immunodetection of the Arabidopsis potassium channel KAT1 (78 kDa, 6 TMDs), in 2D gels. Microsomal proteins from SF9 cells that overexpress KAT1 were used as a starting material. part of it was extracted in 2D buffer (urea + thiourea+ ASB14) and separated on a 2D gel (100 µg loaded). Part of it was extracted in SDS buffer, and loaded in a side lane made on the top of the second dimension gel. Five µg proteins were loaded in this SDS-PAGE lane. The total gel was then blotted and the blot was submitted to immunodetection for KAT1. the signal was quantified and integrated. According to this index, twice as much protein was present in the SDS lane compared to the 2D gel, which means an extraction yield around 2%